\documentclass[lettersize,journal]{IEEEtran}

\usepackage{amssymb,amsmath}
\usepackage{cite}
\usepackage{graphicx}
\usepackage{subfig}

\usepackage{psfrag}
\usepackage{url}
\usepackage[latin1]{inputenc}
\usepackage[absolute,overlay]{textpos}
\usepackage{gensymb}
\usepackage{cases}
\usepackage[font=footnotesize]{caption}
\usepackage{float}
\usepackage[linesnumbered,ruled,lined]{algorithm2e}
\usepackage{pgf}

\usepackage[nocomma]{optidef}

\usepackage{amsthm}

\usepackage{booktabs}
\usepackage{color,soul}

\usepackage{textcomp}
\usepackage{bbm}
\usepackage{xcolor}
\def\BibTeX{{\rm B\kern-.05em{\sc i\kern-.025em b}\kern-.08em
    T\kern-.1667em\lower.7ex\hbox{E}\kern-.125emX}}

\usepackage{tabularx}
\usepackage{makecell}
\usepackage{multirow}

\usepackage{pifont}

\begin{document}
%\title{Explainable Conservative Offline Multi-Agent Learning for 6G Network Slicing}
\title{XAI-Guided Conservative Decentralized Execution for Offline Multi-Agent Network Slicing}
\author{
	\IEEEauthorblockN{Eslam Eldeeb,~\IEEEmembership{Member,~IEEE,}  Hatim~Chergui,~\IEEEmembership{Senior~Member,~IEEE,} and Merouane~Debbah,~\IEEEmembership{Fellow,~IEEE}
}
    \thanks{E. Eldeeb is with the Centre for Wireless Communications (CWC), University of Oulu, Finland. (e-mail: eslam.eldeeb@oulu.fi).
    }
    \thanks{H. Chergui is with the i2CAT Foundation, Spain. (e-mail: chergui@ieee.org)
    }
    \thanks{M. Debbah is with the Research Institute for Digital Future, Khalifa University, 127788 Abu Dhabi, UAE (e-mail: merouane.debbah@ku.ac.ae).}

    \thanks{The work of E. Eldeeb was supported by 6G Flagship (Grant Number 369116) funded by the Research Council of Finland, and supported by the Business Finland project, 6G-FISRE.}
    \thanks{The work of H. Chergui was supported by the grant COALESCE-6G PID2024-163028OB-I00, funded by MICIU/AEI/10.13039/501100011033/FEDER, UE.}
}
\maketitle

\begin{abstract}
The recent advances toward sixth-generation (6G) and beyond-6G networks have accelerated the need for intelligent resource management mechanisms capable of supporting heterogeneous services under shared infrastructures in network slicing. However, resource allocation in network slicing naturally forms a resource-coupled cooperative optimization problem with competing slice demands. Slices compete for limited resources to minimize individual latencies while coordinating to avoid conflicts and underutilization. Although multi-agent reinforcement learning (MARL) has shown promising performance in such settings, existing online formulations remain costly, unsafe, and difficult to deploy due to their reliance on environmental interactions and communication among agents. In this work, we present explainable artificial intelligence (XAI)-guided conservative decentralized execution (X-CODE). X-CODE is an explainable offline MARL that operates offline without environmental interaction, nor inter-agent communication. It exploits explainability-aware reward shaping to modify the relative preference among joint offline transitions during centralized training to improve decentralized resource-allocation behavior. In deployment, the agents operate independently without signaling exchange among the agents. Simulation results demonstrate that the proposed approach achieves zero observed resource-conflict events in the evaluated test episodes while minimizing per-slice latencies. Moreover, the proposed framework exhibits lower signaling overhead and reduces effective inference latency by $88 \%$ under the considered communication-delay model compared to the online baselines. Source codes and datasets are available through:~\url{https://github.com/Eslam211/xcode-ran-slicing}.
\end{abstract}

\vspace{-1mm}

\section{Introduction}\label{sec:introduction}
The recent advances toward 6G and beyond-6G networks have urged the need for continuous control and monitoring of communication systems~\cite{11455199}. Such systems are expected to support a wide range of applications, including enhanced mobile broadband (eMBB), massive machine-type communications (mMTC), and ultra-reliable low-latency communications (URLLC). To enable network coexistence under shared resources, network slicing has been proposed as a key enabler in 6G and beyond-6G systems. Network slicing partitions a physical network into multiple virtual networks (slices), each operating independently to satisfy specific service level agreements (SLAs), such as latency and bandwidth requirements~\cite{11517469,11393619}.

While network slicing demonstrates several benefits, including enhanced security and cost efficiency, resource management still poses several limitations. Key challenges are~\cite{11456226,11124199}:
\begin{enumerate}
    \item emerging applications impose strict and often conflicting requirements that exacerbate resource consumption;

    \item optimizing multiple objectives over a shared physical infrastructure often introduces resource contention;

    \item learning-based policies may converge to conservative resource-allocation strategies that lead to underutilization.
\end{enumerate}
Therefore, resource management in network slicing often introduces conflicting objectives, where each slice competes with other slices for the available resources while simultaneously cooperating to avoid persistent resource overcommitment or underutilization. From optimization point-of-view, resource management in network slicing is seen as a resource-coupled cooperative multi-agent optimization with competing slice demands~\cite{11199352}.

Decentralized multi-agent systems primarily relies on structured communication signals between independent agents to enable constructive coordination. However, it introduces additional complexity and delays due to the high signaling overhead. Meanwhile, it leaves the system prone to failures when communication is limited or unavailable. Similarly, considering a centralized optimization framework, where all agents act as a single entity, is computationally inefficient due to the high dimensionality that increases dramatically with the number of agents~\cite{NEURIPS2025_e26502ce}. 

Recently, reinforcement learning (RL) and multi-agent reinforcement learning (MARL) have shown great success in solving complex single-agent and multi-agent wireless scenarios. Unlike conventional optimization techniques, model-free RL methods can efficiently adapt their policies without accurate mathematical representation of the environment. Online RL learns optimal policies through continuous sequential interactions with the environment. In addition, MARL methods enable distributed decision-making among multiple agents while accounting for the interactions induced by shared network resources~\cite{pmlr-v162-mao22a}.

In MARL, multiple agents learn their optimal policies simultaneously by interacting with both the environment and each other. To overcome the high dimensionality of centralized MARL methods and the high overhead of decentralized MARL, centralized training with decentralized execution (CTDE) was introduced in~\cite{NIPS2017_68a97503}. It enables agents to exploit global information during training, allowing them to learn coordinated behaviors. During execution, each agent relies solely on its local observations and decentralized policies to make decisions independently. Thereby, CTDE eliminates the need for communication among agents during execution. As a result, CTDE provides an effective trade-off between coordination efficiency and execution scalability. However, they still require information sharing during training~\cite{11419150}.

Despite the potential of MARL methods in solving complex multi-agent problems, a key limitation lies in their online formulation. During online CTDE training, agents require continuous exchange of information to enable efficient decentralized policy execution. Such online interaction is often impractical due to long training intervals and high costs. Additionally, online MARL methods pose safety concerns due to the exploration during early training that might lead to unsafe decisions or persistent resource overcommitment~\cite{11314112}. Offline RL/MARL addresses these limitations by shifting the optimization processes to offline settings using static datasets that are collected beforehand. This enables safer and more resource-efficient training without requiring direct interaction with the environment during policy learning. Moreover, it facilitates the deployment of learning-based solutions in real-time, where online training is often undesirable or infeasible~\cite{11103474}.

However, offline MARL methods, especially resource-coupled cooperative MARL with competing demands, remain highly challenging~\cite{Choi_2025}. Offline datasets often exhibit narrow coverage, especially when they are collected using random behavioral policies. Consequently, the learned policies may rely on out-of-distribution (OOD) actions, leading to over-optimistic value estimates and unstable learning behavior~\cite{10753476}. Moreover, the absence of communication among agents further complicates coordination, increasing the likelihood of convergence to sub-optimal or saddle-point solutions. In online MARL, such problems are avoided by collecting additional data through environmental interactions to correct optimistic estimates, which is not feasible in offline MARL. Therefore, implicit coordination among agents is needed during training to stabilize learning under limited data coverage, without requiring explicit online interactions during execution~\cite{10854503,arxiv_NS}.

To address these challenges, in this work, we propose an explainable offline MARL framework for resource-coupled cooperative MARL with competing slice demands. The proposed approach builds upon conservative Q-learning (CQL)~\cite{kumar2020conservative}. CQL is a well-known offline RL algorithm that constrains value estimates by penalizing out-of-distribution (OOD) actions in the offline datasets. Thus, we formulate a multi-agent CQL algorithm based on centralized training with decentralized execution for offline multi-agent CQL to enable decentralized policy execution. To provide additional training guidance under limited dataset quality, we refine the reward signals in the offline dataset using an explainability-guided shaping mechanism. The mechanism uses centralized value attributions to modify the relative preference among joint offline transitions. As demonstrated empirically, this relabeling helps the decentralized policies achieve a more favorable balance between resource utilization and conflict avoidance. Our contributions are summarized as follows:
\begin{itemize}
    \item We formulate the network slicing resource management problem as a resource-coupled cooperative MARL with competing slice demands, where slices compete for limited edge resources while coordinating to avoid resource conflicts or underutilization.

    \item We propose a novel XAI-guided conservative decentralized execution (X-CODE) framework that adopts the CQL algorithm in multi-agent settings using CTDE framework. Moreover, the proposed method leverages explainable AI (XAI) methods (\emph{i.e.}, reward shaping) to provide guidance signals during offline optimization.
    
    \item We provide explainability-driven insights into the coordination process by analyzing SHAP reward-labeling attributions across agents. We analyze SHAP attributions across agents to identify the local features that dominate the centralized value estimates and to interpret the allocation behavior learned by the decentralized policies

    \item Simulation results demonstrate that the proposed approach achieves zero observed resource-conflict events in the evaluated test episodes while minimizing per-slice latencies. Extensive evaluations and ablation studies further highlight the effectiveness of the proposed framework in reducing signaling overhead by $100 \%$ and inference time by $88 \%$ compared to online baselines.
\end{itemize}

The remainder of the paper is organized as follows: Section~\ref{LitRev} reviews related literature. Section~\ref{sec:system_model} introduces the system model and problem formulation. Section~\ref{sec:background} provides the needed preliminaries on online and offline MARL. Section~\ref{sec:x_orl} presents the proposed methodology. Numerical results are provided in Section~\ref{sec:results}, while Section~\ref{sec:conclusions} concludes the paper.

\vspace{-2mm}

\section{Related Work}\label{LitRev}
In this section, we revisit the literature relevant to the paradigm shift from online to offline MARL for network management, following a logical progression that highlights the emergence of trustworthy XAI-augmented RL for 6G systems.
\subsection{Foundational Shifts from Online to Offline MARL in Telecommunications}
Network slicing has emerged as a cornerstone architecture for 6G and beyond systems, enabling operators to partition shared physical infrastructure into multiple isolated logical networks to meet the heterogeneous demands of different services simultaneously~\cite{10.1109/TMC.2023.3328950}. Given the dynamic traffic variations and the complexity of modern radio access networks (RAN), deep reinforcement learning (DRL) has been widely explored to automate resource allocation and optimize slice capacity~\cite{8540003}. However, deploying standard online DRL algorithms in live telecommunication systems introduces severe operational risks. Online agents require continuous, real-time interactions with the environment to learn, which can result in unacceptable latency violations, costly signaling overhead, and severe safety concerns during the exploration phase~\cite{11129914}. Consequently, the focus has shifted toward offline reinforcement learning. Offline RL methods learn effective control policies entirely from static, pre-collected datasets without requiring active environmental interaction, demonstrating a robust ability to maintain service quality and reduce delay violations even when trained on suboptimal historical data~\cite{10632750}.

\subsection{Algorithmic Innovations: Tackling Overestimation and Coordination}
Transitioning to offline data introduces the critical challenge of distributional shift, where evaluating out-of-distribution (OOD) actions leads to extrapolation errors and value overestimation. To counteract this, conservative Q-learning (CQL) was introduced as a foundational offline RL algorithm that applies a regularizer to penalize OOD actions, effectively learning a conservative Q-function that lower-bounds the true policy value~\cite{kumar2020conservative}. Extending these offline architectures to multi-agent reinforcement learning (MARL) settings further exacerbates the overestimation problem due to the exponential explosion of the joint action space~\cite{liu2024offlinemultiagentreinforcementlearning}. To enable cooperative multi-agent learning without demanding prohibitive inter-agent communication during execution, researchers rely heavily on the centralized training with decentralized execution (CTDE) paradigm. Foundational CTDE architectures, such as QMIX, ensure consistency between centralized learning and decentralized policies by enforcing a monotonic factorization of the joint action-value function~\cite{3455716.3455894}. Building on these principles, modern offline MARL frameworks tailor conservative value estimation directly for multi-agent settings. For instance, counterfactual conservative Q-learning (CFCQL) calculates conservative regularizations separately for each agent to prevent the overly pessimistic value estimations that plague direct adaptations of single-agent CQL~\cite{shao2023counterfactual}. Applied directly to telecommunications, these CTDE-based offline MARL frameworks have been successfully formulated to optimize radio resource management (RRM), achieving significant gains in network sum and tail rates while entirely avoiding online interaction overhead~\cite{11463052}.

\subsection{Trustworthy Network Slicing via Explainable AI}
Despite the operational and algorithmic advantages of offline MARL, the inherent \emph{black-box} nature of deep neural networks acts as a significant deterrent to adoption by network operators, particularly in service-level agreement (SLA)-bound or safety-critical 6G environments. To foster operator trust, recent approaches are embedding Explainable Artificial Intelligence (XAI) directly into the reinforcement learning lifecycle. Specifically,~\cite{10283684} introduces a novel integrated architecture where an XAI reward signal is shaped to guide an online multi-agent system through complex state-action spaces without obfuscating the decision rationale. Expanding on this, highly dynamic settings like vehicular network slicing have also seen the integration of feature-attribution techniques. In this respect,~\cite{sun2025explainableaiframeworkdynamic} proposed an interpretable online DRL framework that merges attention mechanisms with Shapley values to explicitly score the contribution of specific network features during resource management. By providing post-hoc explainability and supervising the attention layers with rigorous game-theoretic metrics, these XRL approaches ensure that resource allocation for strict use-cases like Ultra-Reliable Low-Latency Communications (URLLC) is both highly optimal and entirely transparent to network administrators.

Having said that, designing an explanation-guided offline MARL scheme is still unexplored in literature, especially to ensure the required robustness to tackle high stake 6G network slicing scenarios.

\section{Network Setup}\label{sec:system_model}
We consider an edge-enabled RAN slicing system composed of $K=3$ network slices, corresponding to eMBB, URLLC, and mMTC services. The slices share the computing resources of a common edge server with nominal CPU budget $f_{\max}$. We define the default feasible CPU allocation vector as
$\boldsymbol{f}^{\mathrm{th}}=[f^{\mathrm{th}}_1,\ldots,f^{\mathrm{th}}_K]$, where $\sum_{k=1}^{K} f^{\mathrm{th}}_k=f_{\max}$. Let the system operate in discrete decision intervals each has a duration $\tau$. At each decision interval, each slice $k$ selects a CPU allocation $a_{k,t} \in \mathcal{A}$ from a predefined finite set $\mathcal{A}=\{1,5,10,15,20,25,30\}$, where each value denotes a different CPU allocation and is expressed in $10^9$ CPU cycles/s. We consider the traffic arrival during one decision interval as follows:
\begin{equation}
    A_{k,t} = \tau \: \lambda_{k,t}.
\end{equation}
where $\lambda_{k,t}$ and $c_{k,t}$ denote the instantaneous traffic arrival rate and transmission capacity of slice $k$ in bit/s, respectively.

Each slice $k$ is associated with two queues, \emph{i.e.}, a computation queue $q^{\mathrm{c}}_{k,t}$ and a transmission queue $q^{\mathrm{r}}_{k,t}$. Hence, the computation queue evolves according to the CPU allocation as:
\begin{equation}
    q^{\mathrm{c}}_{k,t+1} = \max\left(q^{\mathrm{c}}_{k,t}-S_{k,t},0\right) + A_{k,t},
\end{equation}
where $S_{k,t} = \tau \: U \: a_{k,t}$ is the  number of processed bits during one decision interval given the CPU allocation $a_{k,t}$ and $U$ is the computation efficiency in bit/cycle. Similarly, the transmission queue evolves as follows:
\begin{equation}
    q^{\mathrm{r}}_{k,t+1}= \max\left(q^{\mathrm{r}}_{k,t}-\tau c_{k,t},0\right) + \min\left(q^{\mathrm{c}}_{k,t},S_{k,t}\right).
\end{equation}
Thus, the end-to-end queuing state of slice $k$ depends jointly on the traffic arrival, transmission capacity, and CPU allocation. Since the traffic arrival and transmission capacity are system-dependent, we only control the CPU allocation.

The accumulated queue length of slice $k$ is updated as follows:
\begin{equation}
    Q_{k,t+1} = Q_{k,t} + q^{\mathrm{c}}_{k,t+1} + q^{\mathrm{r}}_{k,t+1}.
\end{equation}
The average end-to-end latency of slice $k$ is computed using Little's law~\cite{little} as follows:
\begin{equation}
L_{k,t+1}=\frac{Q_{k,t+1}}{(t+1)\bar{\lambda}_k}.
\end{equation}
where $\bar{\lambda}_{k}$ is the average arrival rate of slice $k$. Since we consider that all slices share the same edge server, a resource-overcommitment event, referred to hereafter as a resource conflict, occurs when the aggregate requested CPU allocation exceeds the nominal CPU budget as follows:
\begin{equation}\label{conf_eq}
    \chi_t = \mathbbm{1} \left\{\sum_{k=1}^{K} a_{k,t} > f_{\max} \right\},
\end{equation}
where $\mathbbm{1}\{\cdot\}$ is the indicator function. When the aggregate requested CPU exceeds the edge-server budget, the request is declared infeasible and the simulator applies the default feasible allocation $\boldsymbol{f}^{\mathrm{th}}$ instead. Hence, the effective CPU allocation used
in the queue dynamics is given by:
\begin{equation}
\label{effective_action}
    \tilde{a}_{k,t} = (1-\chi_t)a_{k,t} + \chi_t f^{\mathrm{th}}_k,
\end{equation}
where $\chi_t$ is defined in~\eqref{conf_eq}. Therefore, resource conflicts are recorded and penalized, while infeasible CPU allocations are not directly applied
to the queue evolution.

%Although, each slice aims to reduce its own latency, all slices are coupled through the shared CPU constraint. Thus, the actions represent requested rather than guaranteed CPU allocations, and $\chi_t$ measures an overload request at interval $t$.

\subsection{Problem Formulation}
The objective is to jointly allocate the edge-server CPU resources among the $K$ slices such that the individual average latencies are minimized. To minimize the latencies, each slice aims to maximize its resource utilization, which leaves the system prone to conflicts. In contrast, a conservative policy that underutilizes the resources will increase the latencies. Hence, the agents' objective is to find the optimum allocation policies that minimize average latencies, while avoiding excessive use of the shared CPU budget or underutilization. The resource-allocation problem is formulated as follows:
\begin{subequations}\label{P1}
    \begin{alignat}{2}
        \mathbf{P1:}\qquad
        &\underset{\{a_{k,t}\}}{\min}
        &\quad&
        \sum_{t=0}^{T-1}
        \left[
        \sum_{k=1}^{K}\lambda_L L_{k,t+1}
        +
        \lambda_c \chi_t
        \right],
        \label{P1:a}
        \\
        & \: \: \text{s.t.}
        &\quad&
        a_{k,t} \in \mathcal{A},
        \qquad \forall \: k,t.
        \label{P1:b}
    \end{alignat}
\end{subequations}
where $\mathcal{A}=\{1,5,10,15,20,25,30\}$ is the discrete CPU allocation set, $f_{\max}$ is the nominal CPU budget at the edge server, $T$ is the number of elapsed decision intervals, and $\lambda_L$, and $\lambda_c$ are weighting coefficients for latency minimization and conflict avoidance, respectively. The variable $\chi_t$ denotes the resource-conflict indicator, defined in~\eqref{conf_eq}. The CPU budget is handled through the conflict indicator and the default-allocation fallback in~\eqref{conf_eq}; thus, infeasible requests are penalized but the queue dynamics evolve under a feasible applied allocation.

The first term in~\eqref{P1:a} penalizes high slice latencies, while the second term penalizes resource-conflict events in which the aggregate requested CPU exceeds the edge-server capacity. The constraint in~\eqref{P1:b} reflects the available discrete resource-allocation levels supported by the system. The slices are inherently coupled by the same CPU budget, while seeking sufficient resources to reduce their individual latencies. The formulated optimization problem has a resource-coupled cooperative with competing slice structure, where slices compete for limited edge resources, yet coordinated allocation is required to avoid conflicts and maintain stable network-wide performance. Additionally, we consider pre-collected static datasets used to solve the optimization problem without online interaction with the network as further explained in Section~\ref{sec:x_orl}.

\section{Background}\label{sec:background}
\subsection{Online MARL}
Reinforcement learning excels in solving complex model-free problems. In its multi-agent formulation, $K$ agents explore the environment at each time index $t$. Each agent $k$ observes its local observations $o_{k,t}$, takes action $a_{k,t}$, transits to the next observations $o_{k,t+1}$ and receives an immediate reward $r_{k,t}$. The local observations of all agents form the joint observation $\mathbf{o}_t = [o_{1,t}, \dots, o_{K,t}]$, while the action of each agent forms the centralized action $\mathbf{a}_t = [a_{1,t}, \dots, a_{K,t}]$ of the system. We focus on a \emph{partially observable multi-agent reinforcement learning} setting, in which each agent $k$ has limited access only to its local observation $o_{k,t}$ and produces action $a_{k,t}$ by following a \emph{policy} $\pi_k(a_{k,t} \mid o_{k,t})$~\cite{marl-book}.

The desired goal in MARL is to find the optimal policies $\pi^*(a | s) = \{ \pi^*_k(a_k | o_k) \}_{k=1}^{K}$ that maximize the expected return $\Big \{ \mathbb{E} [Z_k^{\pi}] = \mathbb{E}[\sum_{t=0}^\infty \gamma^t r_{k,t}] \Big\}_{k=1}^{K}$, where $\gamma$ is the discount factor. Classical Q-learning methods obtain the optimal policies by finding the optimal Q-functions $\{Q^*_k(o_k,a_k)\}_{k=1}^{K}$ as:
\begin{equation}
\pi_k^*(a_k|o_k)=\mathbbm{1}\left\{a_k =\arg\max_{\tilde a}Q_k^*(o_k,\tilde a)\right\}, \quad k = 1, \dots, K,
\label{policy}
\end{equation}
where $\mathbbm{1}\{\cdot\}$ is the indicator function.

In high dimension problem setting, deep neural networks are used to approximate the Q-function. Deep Q-network (DQN) is a popular algorithm designed for discrete action spaces. DQN iteratively minimizes the Bellman loss for each agent $k$ independently as follows~\cite{bellman1966dynamic,DQNs}:
\begin{align}
\label{DQN_loss}
    \mathcal{L}_k^{\text{DQN}}(\theta_k) = \:& \hat{\mathbb{E}} \left[ \left(r_k +\gamma \max_{\tilde a_k} \hat{Q}_k(o'_k,\tilde a_k) - Q_k(o_k,a_k) \right)^2 \right],
\end{align}
where $\theta_k$ is the learned parameters of the Q-function, $o_k$ is the current observation, $o^{\prime}_k$ is the next observation, $a_k$ is the current action and $a^{\prime}_k$ is the next action. DQN adopts a target Q-network $\hat{Q}_{k}$ to stabilize the performance. Additionally, DQN is an off-policy algorithm, where the parameters are updated using a mini-batch sampled from a replay buffer that has the past experience stored in.

\subsection{Offline MARL via Conservative Q-learning}
In offline MARL, the objective is to find the optimal policies $\{ \pi^*_k(a_k | o_k) \}_{k=1}^{K}$ solely using a static dataset $\mathcal{D}$ without environmental interactions~\cite{levine2020offline}. An offline dataset has the agents' experience $\mathcal{D}=\{(o_{1},\dots, o_{K}, a_{1},\dots, a_{K}, r_{1}, \dots, r_{K}, o^{\prime}_{1}, \dots, o^{\prime}_{K})\}$. Such dataset is usually collected using behavioral policies $\{ \pi^{\beta}_k(a_k | o_k) \}_{k=1}^{K}$ that can be random policies or prior online policies (\emph{e.g.}, replay buffers). A well-known limitation in offline RL is the distributional gap between the behavioral policies and the learned policies, especially when the datasets have limited coverage with sub-optimal action distribution (\emph{e.g.}, random actions). In that case, maximizing over the actions in the TD error in~\eqref{DQN_loss} may yield over-optimistic return estimates due to the out of distribution actions (OOD) seen by the learned policies. In online off-policy RL (\emph{e.g.}, DQN), this problem is resolved by collecting more data, which is not feasible in offline RL~\cite{kostrikov2021offlinereinforcementlearningimplicit}.

\begin{figure*}[t!]
    \centering    \includegraphics[width=1.9\columnwidth,trim={0cm 0cm 0cm 0cm},clip]{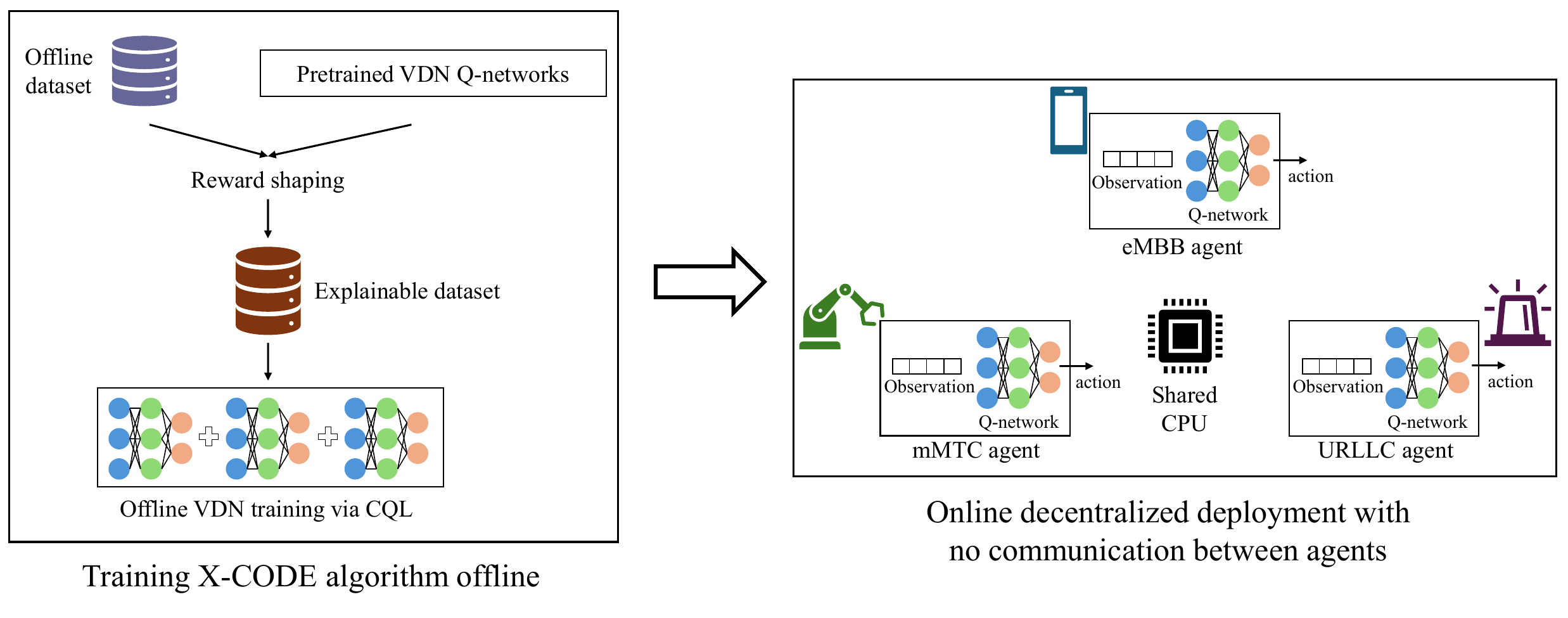} \vspace{0mm}
    \caption{A diagram illustrating the proposed X-CODE Algorithm. An offline dataset and pretrained VDN Q-networks are used to perform reward shaping and construct the explainable dataset. Then, the proposed X-CODE training is performed via value-decomposed CQL-VDN training. Finally, coordinated agents execute their learned policies online without any inter-agent communication.}
    \vspace{0mm}
    \label{X_CODE_Fig}
\end{figure*}

\emph{Conservative Q-learning (CQL)}~\cite{kumar2020conservative} addresses the distributional shift problem by introducing a regularization term in the Bellman update. This term penalizes high Q-values for actions that are not seen in the offline dataset. Hence, the CQL update for each agent $k$ independently is updated as follows:
\begin{align}
\label{CQL_loss}
    \mathcal{L}_k^{\text{CQL}}(\theta_k) = \:& \hat{\mathbb{E}} \left[ \left(r_k +\gamma \max_{a^{\prime}} \hat{Q}_{k}(o_k^{\prime},a_k^{\prime}) 
   - Q_k(o_k,a_k) \right)^2 \right] \\ \nonumber
   & + \alpha \: \hat{\mathbb{E}} \bigg[ \log \sum_{\tilde{a}}
\exp \bigl( Q_k(o_k,\tilde{a}) \bigr) 
    - \ Q_k(o_k,a_k)  \bigg],
\end{align}
where $\alpha$ controls the conservative regularization strength and the expectation $\mathbb{E}$ is taken over samples from the dataset (\emph{i.e.}, $o_k$, $o^{\prime}_k$, $a_k$ $\sim \mathcal{D}$ for each agent $k$).

The CQL regularizer increases the loss when the Q-values of unseen or weakly supported actions become larger than the Q-value of the dataset action. Therefore, it discourages over-optimistic value estimates for OOD actions. In that case, actions seen in the dataset will be penalty free, while actions far from the dataset will be heavily penalized. CQL demonstrates robust performance across diverse domains; however, its performance is still highly driven by the quality of the data making its convergence prone to failure. This challenge is more exacerbating in coordinating multiple-agents offline under limited communication scenarios~\cite{shao2023counterfactual}. The next section introduces the proposed XAI-guided approach for the multi-agent scenario under study in this paper.

\section{XAI-Guided Conservative Decentralized Execution}\label{sec:x_orl}

In this section, we present the proposed XAI-guided conservative decentralized execution (X-CODE) approach. First, we formulate the network slicing problem as a partially-observed Markov decision process (POMDP). Afterwards, we introduce the proposed offline CTDE learning and XAI-guided reward relabeling framework.

\subsection{POMDP Formulation}
As described in Section~\ref{sec:system_model}, we consider a cloud-native platform composed of $K$ agents, where each agent corresponds to a network slice with a running service. At each time step $t$, the agents request CPU allocations according to their local traffic demands targeting jointly minimizing communication latency and resource conflicts. The problem is formulated as a POMDP as follows:
\begin{itemize}
    \item \textbf{Action:} At time step $t$, each agent $k$ chooses a CPU allocation $a_{k,t} \in \mathcal{A}$, where $\mathcal{A}=\{1,5,10,15,20,25,30\}$. The joint action is $\mathbf{a}_t=\left(a_{1,t},\ldots,a_{K,t}\right)$.

    \item \textbf{Observation:} Each agent $k$ observes a local state $o_{k,t}=\left[\Phi_{k,t},\beta_{k,t}, q^{\mathrm{c}}_{k,t}, q^{\mathrm{r}}_{k,t}\right]$, where $\Phi_{k,t}$ is the normalized traffic load and $\beta_{k,t}=f^{\mathrm{th}}_k-\tilde a_{k,t-1}$ is the previous applied-allocation gap. Here, $\tilde a_{k,t-1}$ denotes the effective CPU allocation at the previous decision interval as in~\eqref{effective_action}. For initialization, we set $\tilde a_{k,-1}=f^{\mathrm{th}}_k$ and $\beta_{k,0}=0$. The joint observation is $\mathbf{o}_t=\left(o_{1,t},\ldots,o_{K,t}\right)$.
    
    \item \textbf{Reward:} At each time step $t$, each agent $k$ receives the reward:
    \begin{equation}
        \label{reward}
        r_{k,t}=-\lambda_l L_{k,t+1}+\lambda_a \tilde a_{k,t}-\lambda_c \chi_t ,
    \end{equation}
    where the coefficients $\lambda_l$, $\lambda_a$, and $\lambda_c$ control the latency penalty, CPU allocation incentive, and conflict penalty, respectively. Here, $\chi_t \in \{0,1\}$ is a binary number, where $\chi_t = 1$ in case of conflict and $\chi_t = 0$ otherwise as in~\eqref{conf_eq}. This reward encourages using CPU resources to reduce queuing latency, but penalizes joint allocations that exceed the nominal edge-server CPU budget. Then, we define a team reward as follows:
    \begin{equation}
        \label{team_reward}
        r_t^{\mathrm{team}} = \frac{1}{K} \sum_{k=1}^K r_{k,t}.
    \end{equation}
    Although the environment has mixed cooperative-competitive structure through individual latency rewards and a shared conflict penalty, we construct a common team return that combines the slice-level rewards to be used at the learning level as explained next.
\end{itemize}

\subsection{The Proposed X-CODE Algorithm}

The formulated POMDP can be solved using conventional MARL methods. However, offline multi-agent learning is challenging under limited dataset coverage, especially in offline settings under low-quality behavior and missing coordination datasets (\emph{e.g.}, random) and without explicit communication during execution. Hence, we propose X-CODE, illustrated in Fig.~\ref{X_CODE_Fig}, based on the centralized training decentralized execution (CTDE) framework~\cite{NIPS2017_68a97503}. In CTDE, agents can exploit centralized information during training, while each agent executes its policy independently using only its local observation~\cite{amato2024introductioncentralizedtrainingdecentralized}. For compactness, we omit the time index in the following learning formulation and denote a sampled offline transition as
\[
(\mathbf{o},\mathbf{a},r^{\mathrm{team}},\mathbf{o}^{\prime})\in\mathcal{D},
\]
where $\mathbf{o}^{\prime}$ is the next joint observation. 

We propose XAI-guided conservative decentralized execution (X-CODE) framework that relies on offline centralized training and implicit coordination between agents while ensuring effective decentralized policy extraction during deployment without inter-agent communication. X-CODE estimates the centralized action-value function using value decomposition networks (VDN) as follows~\cite{sunehag2017valuedecompositionnetworkscooperativemultiagent}:
\begin{equation}
\label{value_dec_eq}
   Q(\mathbf{o},\mathbf{a})=\sum_{k=1}^{K}\tilde{Q}_k(o_k,a_k),
\end{equation}
where $\tilde{Q}_k(o_k,a_k)$ represents the contribution of agent $k$ to the centralized Q-function.

Hence, the CQL-VDN loss function is formulated as follows:
\begin{align}
\label{CQL_CTDE}
\mathcal{L}_{\mathrm{VDN}}=\hat{\mathbb{E}}\left[\left(y_{\mathrm{VDN}}-\sum_{k=1}^{K}\tilde{Q}_k(o_k,a_k)\right)^2+\alpha \mathcal{R}_{\mathrm{CQL}}\right],
\end{align}
where the first term is the value-decomposition Bellman error under the team reward. In~\eqref{CQL_CTDE}, $y_{\mathrm{VDN}}$ is the TD target calculated as follows:
\begin{align}
y_{\mathrm{VDN}}=r^{\mathrm{team}}+\gamma\sum_{k=1}^{K}\max_{\tilde a_k}\hat{Q}_k(o'_k,\tilde a_k),
\label{eq:ctde_target}
\end{align}
where $\hat{Q}_k$ denotes the target Q-network and $\gamma$ is the discount factor. The second term $\mathcal{R}_{\mathrm{CQL}}$ in~\eqref{CQL_CTDE} is the CQL loss that penalizes high values assigned to actions not supported by the offline dataset and calculated as follows:
\begin{align}
\mathcal{R}_{\mathrm{CQL}}= \frac{1}{K}\sum_{k=1}^{K}\left[\log\sum_{\tilde a_k}\exp\left(\tilde{Q}_k(o_k,\tilde a_k)\right)-\tilde{Q}_k(o_k,a_k)\right],
\label{eq:cql_regularizer}
\end{align}
$\alpha$ controls the conservative regularization strength. After training, each agent extracts its decentralized policy as:
\begin{equation}
\label{Policy_extraction}
\pi_k(o_k)=\underset{a_k\in\mathcal{A}}{\arg\max}\;\tilde{Q}_k(o_k,a_k).
\end{equation}
Thus, centralized reasoning is used only during offline training, while no inter-agent communication is required during deployment.

The key advantage of the formulated CQL-VDN lies in its decentralized execution. As noted in the policy extraction in~\eqref{Policy_extraction}, the policy of each agent is extracted using the decentralized Q-functions and does not require any communication with other agents. While being effective in cooperative settings, VDN remains challenging in resource-coupled cooperative with competing demands settings, where conflicting objectives can complicate agent coordination and stable policy learning~\cite{shao2023counterfactual}. For example, a common sub-optimal policy in network slicing is the conservative policy (\emph{i.e.}, resource underutilization), where each agent underallocates its CPU resources to completely avoid conflicts. Consequently, limited data coverage in the offline setting and the absence of explicit coordination signals can further degrade the learned policies.

To further improve offline coordination, X-CODE performs XAI-guided reward relabeling. We adopt a cross-fitted relabeling procedure, where the offline dataset is partitioned into $M$ disjoint folds $\{\mathcal{D}^{(m)}\}_{m=1}^{M}$. For each fold $m$, a preliminary CQL-VDN critic $\{\tilde Q_k^{\mathrm{pre},(m)}\}_{k=1}^{K}$ is trained using $\mathcal{D}\setminus\mathcal{D}^{(m)}$ via~\eqref{CQL_CTDE}. Hence, for a held-out transition $(\mathbf{o},\mathbf{a},r^{\mathrm{team}},\mathbf{o}')$, the cross-fitted centralized decomposed value is:
\begin{equation}
Q_{\mathrm{VDN}}^{\mathrm{pre},(m)}(\mathbf{o},\mathbf{a})
=
\sum_{k=1}^{K}\tilde Q_k^{\mathrm{pre},(m)}(o_k,a_k).
\end{equation}
The dataset partitioning is essential to reduce bias from explaining and relabeling the same samples used to fit the critic. Afterwards, SHAP values are computed only for the held-out transitions in $\mathcal{D}^{(m)}$.

Let $\alpha_{j,k}$ denote the SHAP value of feature $j$ of agent $k$ with respect to
$Q_{\mathrm{VDN}}^{\mathrm{pre},(m)}(\mathbf{o},\mathbf{a})$, where $j\in\{1,\ldots,d_o\}$ and $d_o$ is the local observation dimension~\cite{lundberg2017unifiedapproachinterpretingmodel}. We map the absolute SHAP values into a feature-attribution distribution:
\begin{equation}
p_{j,k}=\frac{\exp\left(|\alpha_{j,k}|\right)}{\sum_{j'=1}^{d_o}\exp\left(|\alpha_{j',k}|\right)}.
\label{eq:xai_prob}
\end{equation}
The attribution entropy of agent $k$ is computed as
\begin{equation}
H_k=-\sum_{j=1}^{d_o}p_{j,k}\log p_{j,k}.
\label{eq:xai_entropy}
\end{equation}
A low entropy indicates that the attribution mass is concentrated on a small subset of dominant features, while a high entropy indicates a more diffuse and less concentrated attribution profile. Therefore, we interpret the resulting score as an attribution-concentration score.

%{\LinesNumberedHidden
\begin{algorithm}[!t]
\SetAlgoLined

\textbf{Input:} Offline dataset $\mathcal{D}$, number of folds $M$, discount factor $\gamma$,
conservative penalty $\alpha$, shaping strength $\mu$, and number of agents $K$

Partition $\mathcal{D}$ into $M$ disjoint folds $\{\mathcal{D}^{(m)}\}_{m=1}^{M}$

\For{$m=1,\ldots,M$}{
Train preliminary critics $\{\tilde Q_k^{\mathrm{pre},(m)}\}_{k=1}^{K}$ on
$\mathcal{D}\setminus\mathcal{D}^{(m)}$ using $\mathcal{L}_{\mathrm{VDN}}$~\eqref{CQL_CTDE}

\For{\text{each transition in $\mathcal{D}^{(m)}$}}{
Compute SHAP values $\{\alpha_{j,k}\}$ for
$Q_{\mathrm{VDN}}^{\mathrm{pre},(m)}(\mathbf{o},\mathbf{a})$

Compute $p_{j,k}$ and $H_k$ using~\eqref{eq:xai_prob},~\eqref{eq:xai_entropy}

Compute $b_{\mathrm{VDN}}(\mathbf{o},\mathbf{a})$ using~\eqref{eq:ctde_xai_bonus}
}
}

Relabel each transition reward using~\eqref{eq:xai_reward_ctde}

Construct $\mathcal{D}^{\mathrm{XAI}}$ by replacing $r^{\mathrm{team}}$ with $r^{\mathrm{XAI}}$

\While{\text{not converged}}{
Sample a batch from $\mathcal{D}^{\mathrm{XAI}}$

Update the critics using $\mathcal{L}_{\mathrm{VDN}}$ in~\eqref{CQL_CTDE}
}

\textbf{Return:} Optimized decentralized critics $\{\tilde{Q}_k(o_k,a_k)\}_{k=1}^{K}$

\caption{The proposed XAI-guided conservative decentralized execution (X-CODE) algorithm}
\label{CCQL_CTDE_Algorithm}
\end{algorithm}
%}

Since the VDN critic is optimized through a centralized decomposed value, we aggregate the attribution-concentration scores across agents:
\begin{equation}
b_{\mathrm{VDN}}(\mathbf{o},\mathbf{a})=\frac{1}{K}\sum_{k=1}^{K}\left(1-\frac{H_k}{\log d_o}\right),
\label{eq:ctde_xai_bonus}
\end{equation}
where $\log d_o$ normalizes the entropy, and
$1-H_k/\log d_o$ measures the concentration of the attribution profile of agent $k$. A high value indicates that the critic output is primarily associated with a small subset of features. Hence, $b_{\mathrm{VDN}}(\mathbf{o},\mathbf{a})$ increases when the decomposed critics exhibit concentrated feature attributions across agents. Finally, the offline reward is relabeled as:
\begin{equation}
r^{\mathrm{XAI}}=r^{\mathrm{team}}+\mu\left(b_{\mathrm{VDN}}(\mathbf{o},\mathbf{a})-\bar{b}_{\mathrm{VDN}}\right),
    \label{eq:xai_reward_ctde}
\end{equation}
where $\mu$ controls the shaping strength and $\bar{b}_{\mathrm{VDN}}$ is the empirical average of the bonus over the offline dataset:
\begin{equation}
    \bar{b}_{\mathrm{VDN}}=\frac{1}{|\mathcal{D}|}\sum_{(\mathbf{o},\mathbf{a},\cdot,\mathbf{o}^{\prime})\in\mathcal{D}}b_{\mathrm{VDN}}(\mathbf{o},\mathbf{a}).
    \label{eq:mean_ctde_bonus}
\end{equation}

The centering term preserves the empirical average reward scale, while modifying the relative preference among offline transitions~\cite{10283684}. Since the same shaping term is applied to the team reward, the relabeling ranks joint transitions at the system level rather than providing individualized agent-specific credit assignment. The resulting dataset $\mathcal{D}^{\mathrm{XAI}}$ is constructed by replacing $r^{\mathrm{team}}$ with $r^{\mathrm{XAI}}$. The CQL-VDN objective in~\eqref{CQL_CTDE} is then retrained using $\mathcal{D}^{\mathrm{XAI}}$. Algorithm~\ref{CCQL_CTDE_Algorithm} summarizes the proposed X-CODE procedure. 

It is important to note that the proposed relabeling is not potential-based reward shaping and therefore does not generally preserve the optimal policy of the original environment reward. Centering $b_{\mathrm{VDN}}$ preserves its empirical mean over the offline dataset, but it does not guarantee policy invariance. Accordingly, $\mu$ controls the trade-off between optimizing the original latency-conflict objective and favoring transitions with concentrated critic attributions, as further studied in the ablation results. The $\mu=0$ case corresponds to optimization of the original reward without XAI-based relabeling. Since the agents must avoid conflicts while competing for shared resources, the proposed shaping signals provide additional guidance regarding the contribution of local features to the centralized Q-function. Hence, the relabeled reward biases offline learning toward transitions whose centralized critic exhibits more concentrated and interpretable feature-attribution patterns. The resulting decentralized policies are empirically evaluated in terms of conflict avoidance, latency, and resource utilization without requiring explicit communication during execution. During inference, each agent relies on its local Q-function to extract its policy as in~\eqref{Policy_extraction}.

\section{Experimental Analysis}\label{sec:results}

\begin{figure*}[t!]
    \centering
    \subfloat[Per-step reward\label{online_rew}]{\includegraphics[width=0.6875\columnwidth]{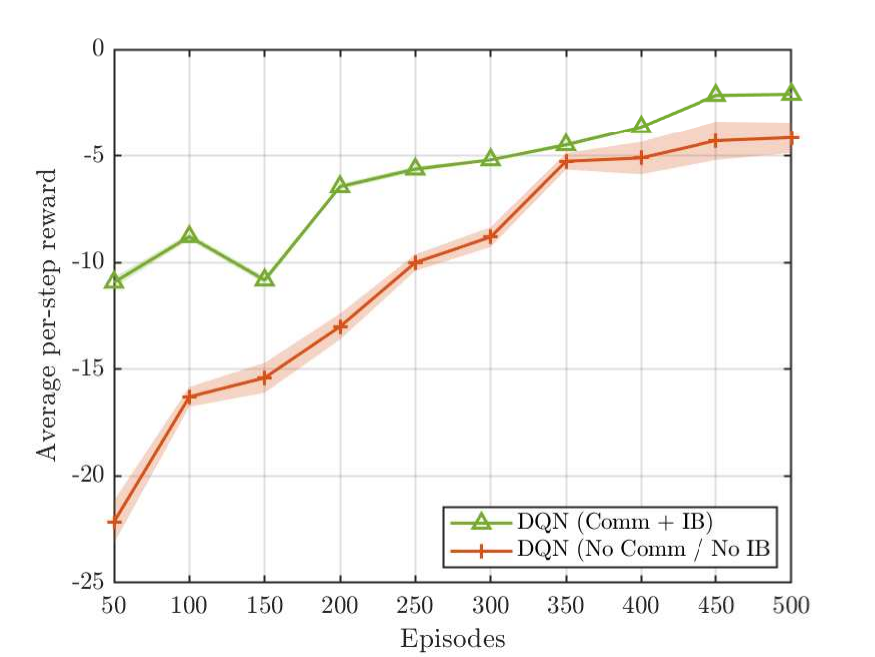}}
    %\vspace{-1mm}
    \subfloat[Average latency\label{online_lat}]{\includegraphics[width=0.6875\columnwidth]{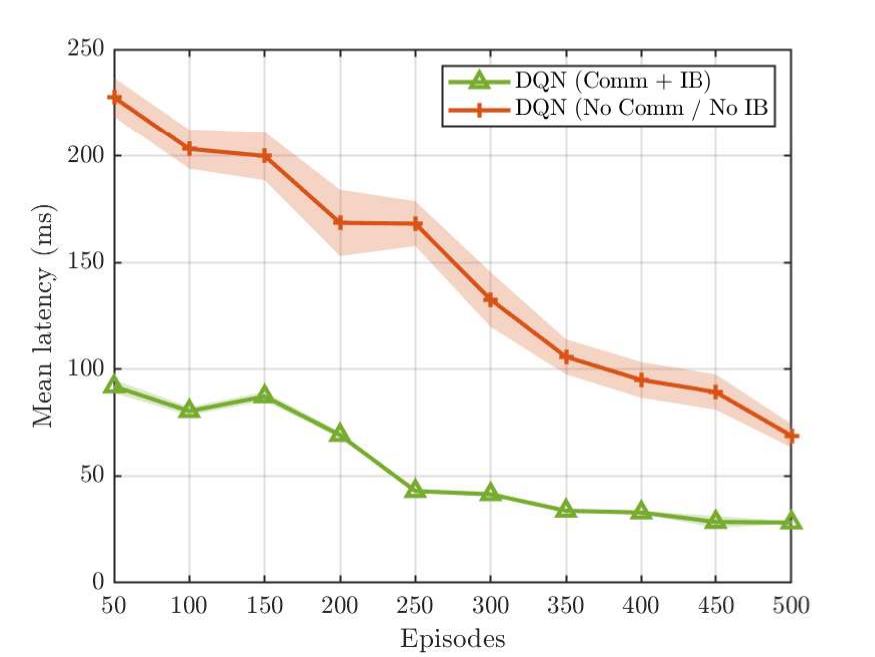}}
    %\vspace{-1mm}
    \subfloat[Conflict rate\label{online_conf}]{\includegraphics[width=0.6875\columnwidth]{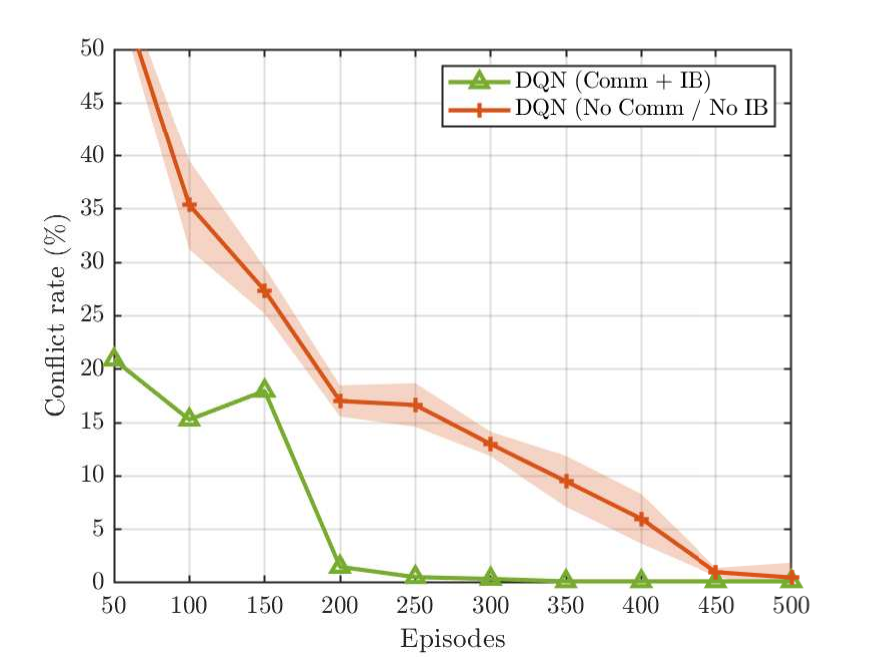}}
    \vspace{-1mm}
    \caption{Performance of online training compared to the heuristic baseline in terms of: (a) per-step reward, (b) average latency, and (c) conflict rate.}
    \vspace{-2mm}
    \label{Online_perf}
\end{figure*}

\subsection{Baselines and Setup}
\subsubsection{Baselines}\label{baselines_sec}
We compare the performance of the proposed X-CODE to several baselines as follows:
\begin{itemize}
    \item \textbf{DQN (No Comm / No IB):} We consider the online baseline DQN (No Comm / No IB), where independent DQNs are trained without any communication between the agents.

    \item \textbf{DQN (Comm + IB):} Following~\cite{10702574}, we consider online DQN (Comm + IB), which is an emergent communication protocol that relies on communication codes between agents and an information bottleneck (IB) module. At every time step $t$, each agent transmits a communication code $m_{k,t} \in \mathcal{M} = \{0,1,2\}$ and receives a set of messages $\bar{m}_{k,t}$ from other agents. The IB module takes as an input the local observation and the communication messages $(o_{k,t}, \bar{m}_{k,t})$ and produces a latent space $x_{k,t}$ that extracts the relevant information and removes redundancies~\cite{7133169}:
    \begin{equation}
        x_{k,t} = \text{IB-AE} (o_{k,t}, \bar{m}_{k,t}),
    \end{equation}
    where $\text{IB-AE} (\cdot)$ is the encoder. Hence, the DQN produces two actions, \emph{i.e.}, main action $a_{k,t}$ and the communication message $m_{k,t}$. The online DQN (Comm + IB) has shown superior performance in such scenarios and reaches near optimal performance online~\cite{10702574}. 

    \item \textbf{XAI-guided independent CQL (X-CQL)} We consider training independent agents as in~\eqref{CQL_loss} while relabeling independent rewards for each agent using local Q-functions.
    
    \item \textbf{Counterfactual conservative Q-learning (CFCQL):} We consider CFCQL~\cite{shao2023counterfactual} as an offline MARL baseline known for its strong performance. CFCQL calculates conservative regularizations separately for each agent to prevent the overly pessimistic value estimations. To ensure fair comparison, we build the CFCQL algorithm on top of the same CQL architecture presented in our approach.

    \item \textbf{QMIX:} We consider the QMIX~\cite{3455716.3455894} algorithm as a strong CTDE baseline compared to the deployed X-CODE. To ensure fair comparison, we unify the simulation parameters between QMIX and VDN on top of the same CQL architecture.

    \item \textbf{VDN:} We consider the VDN~\cite{sunehag2017valuedecompositionnetworkscooperativemultiagent} algorithm as a CTDE baseline compared to X-CODE. Here, VDN corresponds to directly training the objective in~\eqref{CQL_CTDE} without relabeling the reward as in X-CODE.    
\end{itemize}

In addition to these baselines, we perform ablation studies on the main components of the proposed method, including hyperparameters sweeping. We compare each method using several metrics, such as latency and conflict cumulative density function (CDF), training rewards, and CPU utilization. For better visualization and evaluation of different methods, we only show the effective part of the latency CDF rather than the full distribution. We also provide deeper insights on the kernel density estimation (KDE) of the SHAP values.

\begin{table}[h]
\centering
\caption{Simulation parameters and hyperparameters}
\label{Parameters}
\begin{tabular}{cc|cc}
\toprule
\textbf{Parameter}                                    & \textbf{Value} & \textbf{Parameter}                                    & \textbf{Value} \\ \midrule
\midrule

$K$ & $3$ & $T$ & $40$ \\
$f_{\text{max}}$ & $40$ & $f_{\mathrm{th}}$ & $[15,15,10]$ Gcycles/s \\
$\tau$ & $0.01$ s & $U$ & $5 \times 10^{-3}$ bits/cycle \\
$\alpha$ & $1$ & $\mu$ & $1$ \\
$\lambda_c$ & $30$ & $\lambda_l$ & $150$ \\
$\lambda_a$ & $0.05$ & Replay buffer & $10,000$ \\
Optimizer & Adam & Target network coeff. & $0.005$ \\
Hidden layers & $3$ & Neurons & $128$ \\
Online LR & $10^{-3}$ & Offline LR & $10^{-4}$ \\
Online Batch & $64$ & Offline Batch & $256$ \\
Online episodes & $500$ & Offline steps & $1500$ \\
\bottomrule
\end{tabular} 
\end{table}

\begin{figure*}[t!]
    \centering
    \subfloat[Agent $0$\label{data_agent0}]{\includegraphics[width=0.6875\columnwidth]{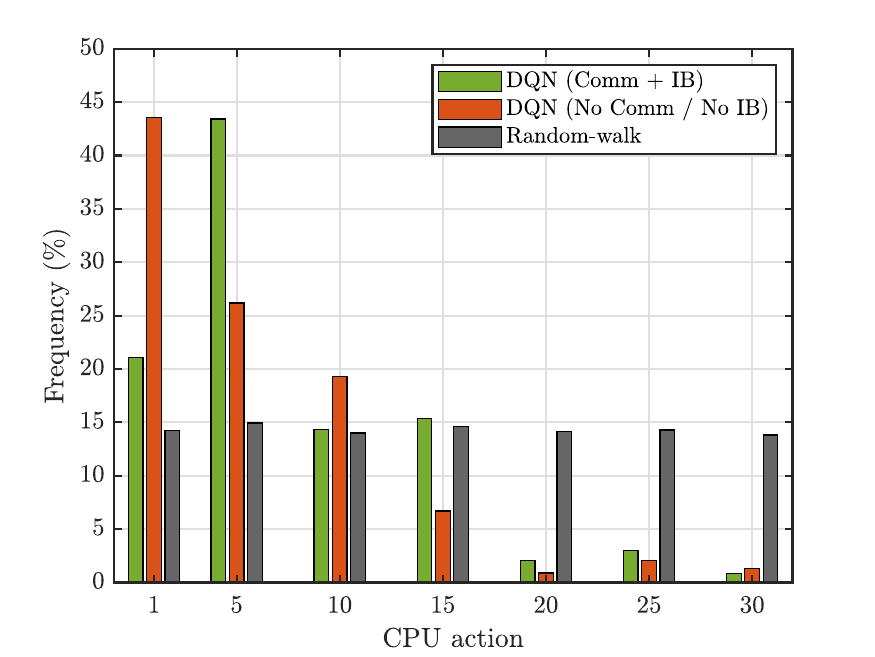}}
    %\vspace{-1mm}
    \subfloat[Agent $1$\label{data_agent1}]{\includegraphics[width=0.6875\columnwidth]{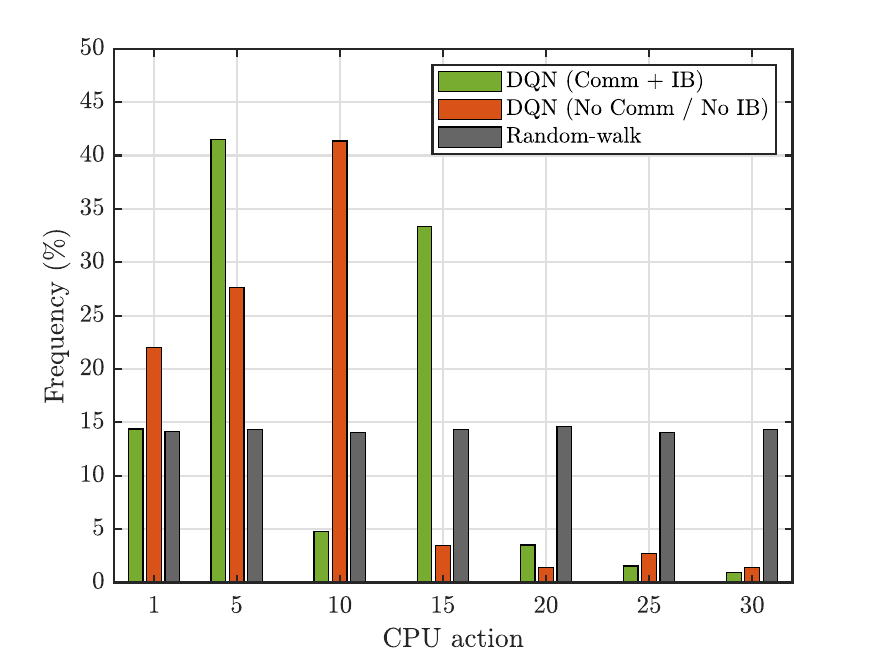}}
    %\vspace{-1mm}
    \subfloat[Agent $2$\label{data_agent2}]{\includegraphics[width=0.6875\columnwidth]{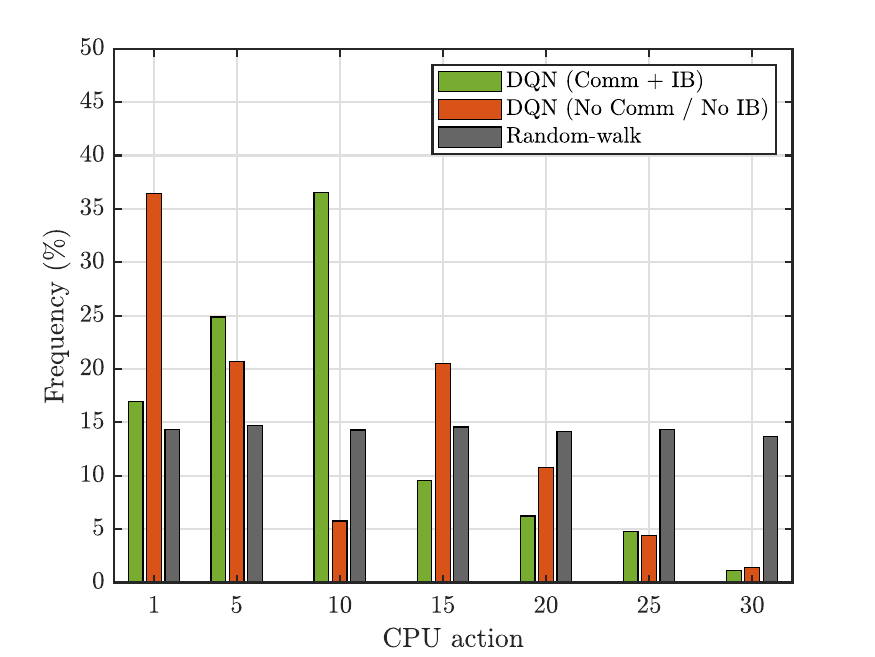}}
    \vspace{-1mm}
    \caption{A comparison of the action coverage of datasets collected during training the online baselines compared to the uniform random-walk dataset.}
    \vspace{-2mm}
    \label{Dataset_coverage}
\end{figure*}

\subsubsection{Setup}
We consider the simulation parameters shown in Table~\ref{Parameters}. We simulate an environment composed of $3$ agents / slices, namely, eMBB (agent $0$), mMTC (agent $1$) and URLLC (agent $2$). Each agent is associated with a dedicated service and aims to allocate CPU resources according to their traffic profiles. Their target is to compete for resources to minimize their individual latencies while cooperating to avoid conflicts. We consider an episodic environment that consists of discrete time steps $t= 1, \dots, T$, where an episode terminates when it reaches a predefined step $T$. We simulate the traffic profiles of each agent according to~\cite{9933014}. The total available resources at the edge server is $40$ Giga cycles / s. For the reward function coefficients in~\eqref{reward}, we set $\lambda_c = 30$, $\lambda_l = 150$, and $\lambda_a = 0.05$. We perform state and reward normalization to ensure stable training. 

For all deployed methods, we model the Q-function using fully-connected neural networks with $3$ hidden layers, each with $128$ neurons. We soft update the target networks with a coefficient $0.005$. For the online baselines, we train each method for $500$ episodes using mini-batch of size $64$, a learning rate (LR) of $10^{-3}$, and Adam optimizer. We set the discount factor $\gamma$ to $0.99$ and use a replay buffer of size $10,000$. Similar to~\cite{10702574}, we adapt $\epsilon$-decay for exploration and prioritized replay buffer. For the proposed X-CODE, we set the LR to $10^{-4}$, batch size to $256$, the cql parameter $\alpha = 1$ and the shaping strength value $\mu = 1$. We collect an offline dataset, whose size is $10,000$, using a random-walk behavioral policy, where all actions are selected randomly. We train all offline methods for $1500$ gradient steps.

\begin{figure*}[!t]
    \centering
    \subfloat[Latency \label{lat_cdf}]{\includegraphics[width=0.6875\columnwidth]{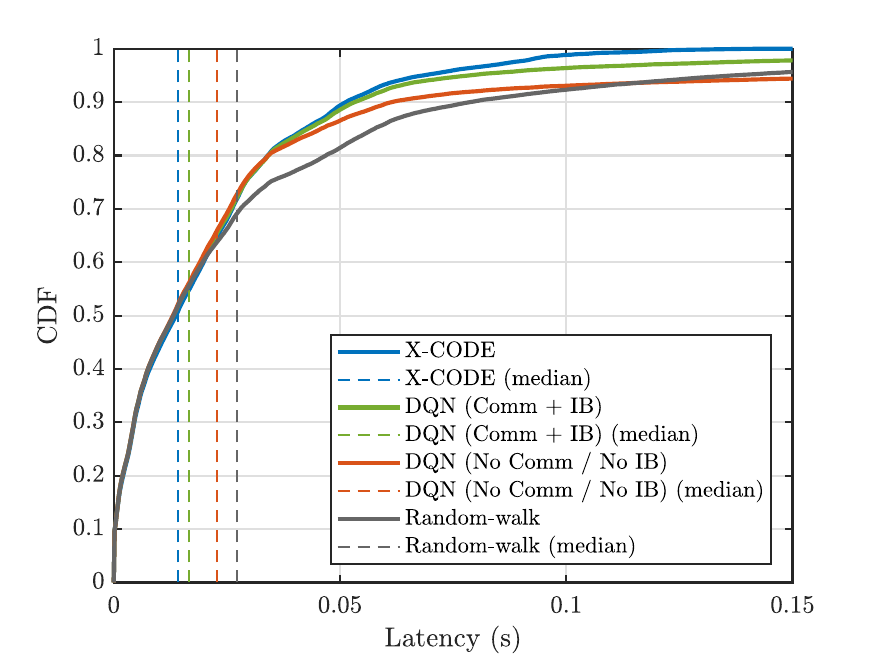}}
    \subfloat[Conflict \label{conf_cdf}]{\includegraphics[width=0.6875\columnwidth]{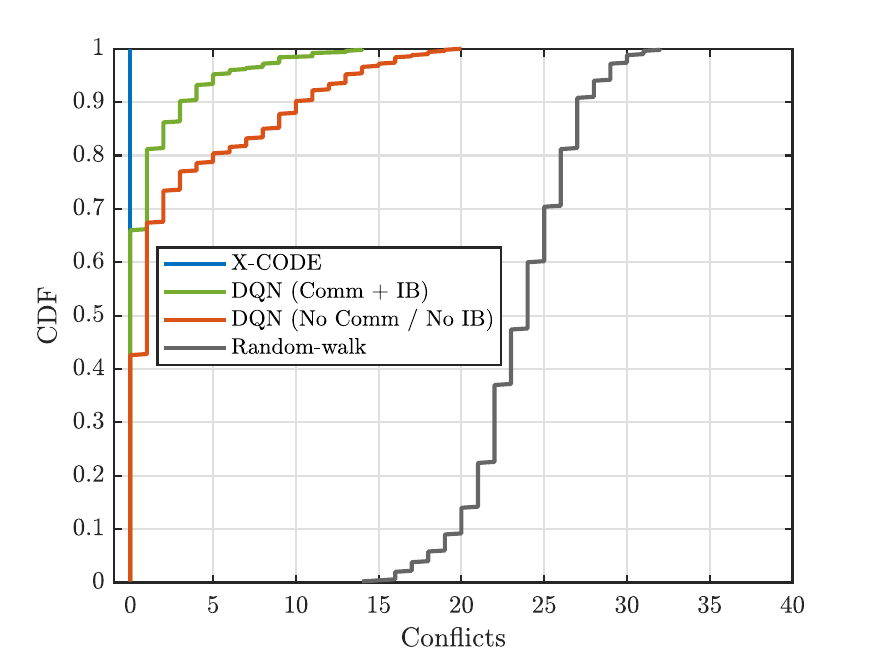}}
    %\hskip -2.8ex
    \caption{The performance of the proposed X-CODE compared to the baselines in terms of the latency and conflict CDFs, respectively.}
    \label{Main_cdfs} %\vspace{-0mm}
\end{figure*}

\subsection{Online Baselines and Datasets}

In the first experiment, we demonstrate the performance of the baselines in the online setting. The objective of this experiment is to show the importance of communication between agents and online interactions in the formulated problem and to use its outcomes as a strong baseline. Fig.~\ref{Online_perf} compares the online learning behavior of the baseline DQN with emergent communication and information bottleneck, \emph{i.e.}, DQN (Comm + IB) against the baseline DQN without communication or IB, \emph{i.e.}, DQN (No Comm / No IB). Overall, both methods improve as training progresses. However, the DQN (Comm + IB) converges faster and achieves better performance across all metrics relying on the communication codes between the agents. In Fig.~\ref{online_rew}, the DQN (Comm + IB) method obtains consistently higher per-step reward throughout training. This indicates that communication and IB help the agents learn more effective resource-allocation decisions earlier in training.

Fig.~\ref{online_lat} further shows that DQN (Comm + IB) significantly reduces the mean latency compared to the DQN (No Comm / No IB) method. After $500$ training episodes, DQN (Comm + IB) reaches less than $30$ ms mean latency, while the DQN (No Comm / No IB) remains above $60$ ms. This demonstrates the benefit of online coordination messages among slice agents to enhance per-agent latencies. Similarly, Fig.~\ref{online_conf} shows that DQN (Comm + IB) eliminates resource conflicts after $200$ episodes, while DQN (No Comm / No IB) requires substantially more training episodes to reach a low conflict rate. While DQN (No Comm / No IB) successfully achieves the target cooperative objective (\emph{i.e.}, avoid conflicts), it underperforms in terms of the competitive objectives (\emph{i.e.}, minimize individual latencies). In contrast, DQN (Comm + IB) excels in both objectives. Overall, these results show that emergent communication improves coordination under the shared CPU constraint, while jointly minimizing per-agent latencies and system conflicts.

Fig.~\ref{Dataset_coverage} compares the action coverage of datasets generated by the online baselines against the random-walk dataset. The objective of this experiment is to provide a comprehensive overview about the quality and coverage of the collected offline datasets prior to offline training. As shown, expert datasets collected from the online policies are more concentrated around specific CPU allocations related to the optimum policies. Hence, these distributions show that the online policies do not explore the action space uniformly; however, they produce rich datasets that contain many optimal learned resource-allocation behavior of each agent. Such datasets generally facilitate offline learning by providing higher-quality samples and better action coverage around desirable operating regions. In reality, expert datasets rarely exist without prior online training or strong behavioral approaches.

In contrast, random-walk dataset provides nearly uniform coverage over the CPU action space for all agents, with each action appearing with approximately similar probability. While the random-walk provides equal action-space exploration, it lacks encoding meaningful coordination or latency-aware behavior. This makes the offline multi-agent optimization even harder and increase the likelihood of convergence to sub-optimal policies. Throughout the remainder of the experiments, we rely on datasets collected using random-walk behavioral policy for the offline training. The objective is to demonstrate the effectiveness of the proposed offline schemes to converge to the optimum policy using low-quality datasets.

\begin{figure}[!t]
    \centering
    \subfloat[CDF \label{CDF}]{\includegraphics[width=0.6875\columnwidth]{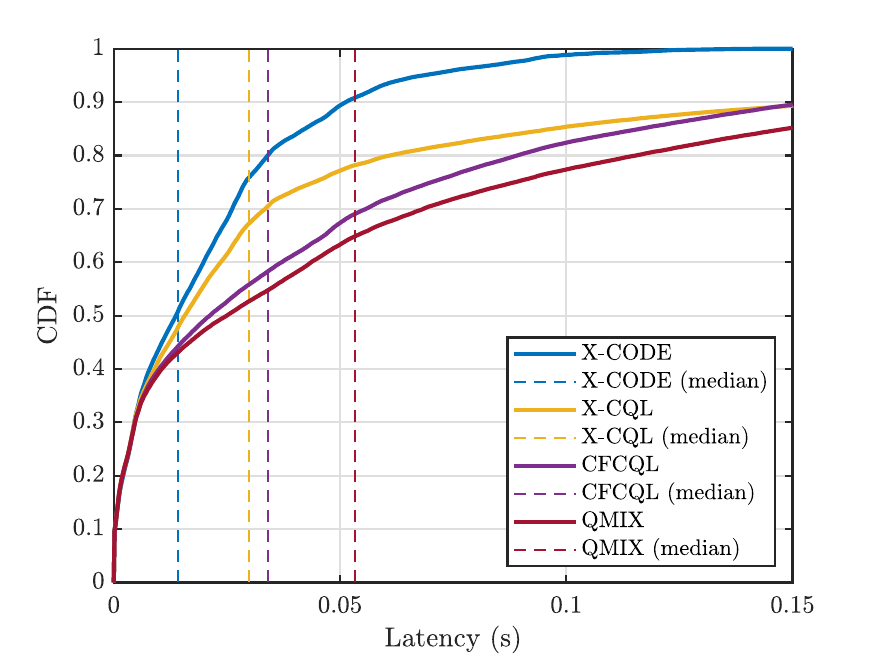}}\\
    \subfloat[CPU utilization \label{cpu}]{\includegraphics[width=0.6875\columnwidth]{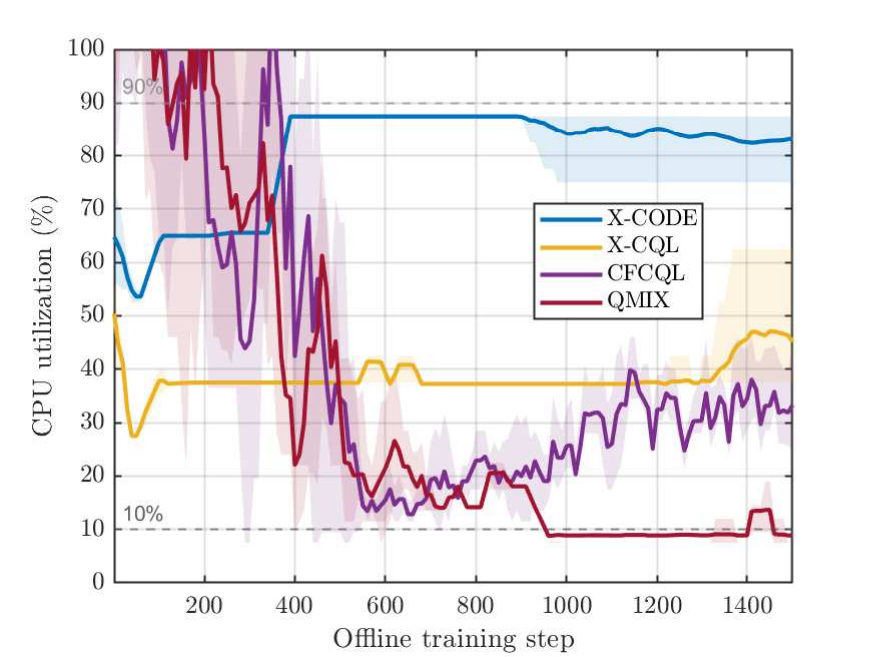}}
    %\hskip -2.8ex
    \caption{The performance of the proposed X-CODE compared to MARL baselines in terms of the latency CDF and the CPU utilization.}
    \label{Ablation_CTDE} %\vspace{-0mm}
\end{figure}

\subsection{Offline Performance}

In the second experiment, we evaluate the performance of the proposed offline X-CODE approach and compare it with the online baselines and the random-walk policy. Fig.~\ref{Main_cdfs} reports the CDFs of the latency and conflict distributions. The objective of this experiment is to assess whether the proposed offline framework can learn an effective slicing policy from a low-quality random dataset while maintaining low latency and avoiding resource conflicts. Fig.~\ref{lat_cdf} shows that X-CODE achieves similar latency distribution to the online DQN (Comm + IB), where both median latencies are almost aligned. It also outperforms other baselines, \emph{i.e.}, DQN (No Comm / No IB) and random-walk.

Fig.~\ref{conf_cdf} further demonstrates that the proposed X-CODE approach provides the strongest conflict-avoidance behavior among all methods. It is the only approach that achieves zero observed resource-conflict events, \emph{i.e.}, no joint allocation exceeded the nominal CPU budget during the evaluated test episodes. In contrast, both online baselines and the random-walk suffer from relatively high conflicts across the slices. Overall, we can observe several benefits from Fig.~\ref{Main_cdfs}. First, the proposed X-CODE shifts the optimization completely to be offline, which demonstrates safe and resource-preserving training. Second, it mitigates communication between agents during inference, relying on decentralized critics during execution. Consequently, this significantly reduces high signaling overhead and enhances inference time needed to successfully execute the policies. Finally, it balances the trade-off between competitive and cooperative objectives by minimizing the per-agent latencies while avoiding conflicts under the evaluated traffic conditions.

Fig.~\ref{Ablation_CTDE} compares X-CODE with other MARL baselines as explained in Section~\ref{baselines_sec}. The objective of this experiment is to investigate whether centralized training and reward shaping improves offline coordination under the shared CPU constraint against competitive MARL methods. As shown in Fig.~\ref{CDF}, QMIX records the worst latency CDF, while X-CQL and CFCQL achieves relatively better latency distribution. The proposed X-CODE achieves a significantly better latency distribution than all baselines with a relatively large median latency gap. This indicates that centralized value decomposition and reward shaping assist the agents to learn more guided resource-allocation decisions from the offline dataset. The baselines exhibit heavier latency tails, which shows that the baselines are less effective in capturing the coupling induced by the shared CPU budget.

Fig.~\ref{cpu} illustrates the CPU utilization during offline training as function of the training steps. Both CFCQL and QMIX aggressively avoid conflicts by converging to conservative policies that underutilize the available resources. Similarly, X-CQL remains significantly below the high-utilization threshold for most of the training process, reflecting a more conservative policy that underutilizes the shared CPU resources. In contrast, the proposed X-CODE progressively increases CPU utilization with more offline steps until converging toward high-utilization region, while avoiding allocations above the nominal resource budget and effectively mitigates conflicts (as reported in Fig.~\ref{conf_cdf}).

\begin{figure}[!t]
    \centering
    \subfloat[CQL parameter $\alpha$ \label{alpha}]{\includegraphics[width=0.6875\columnwidth]{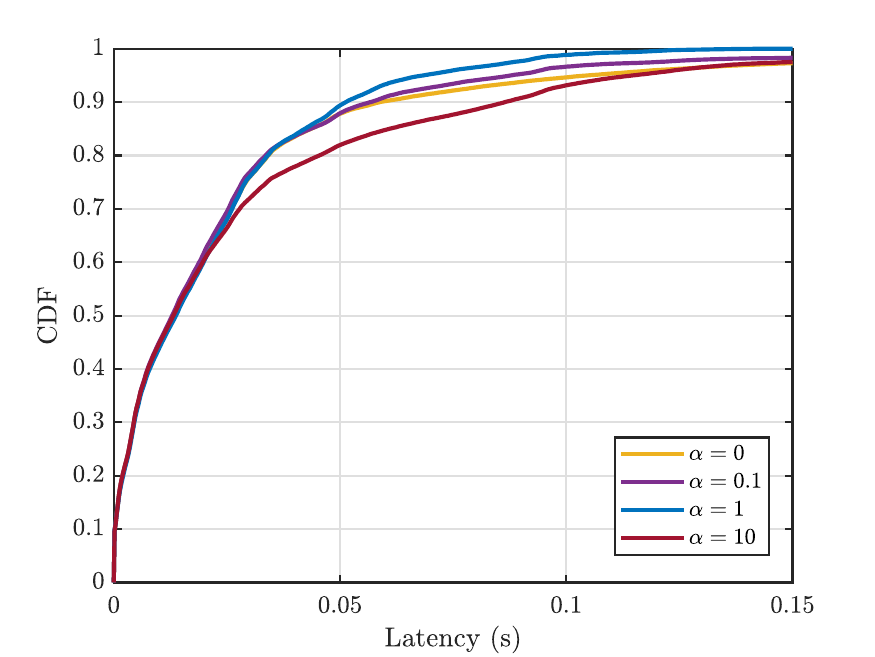}}\\
    \subfloat[Explainability strength $\mu$ \label{mu}]{\includegraphics[width=0.6875\columnwidth]{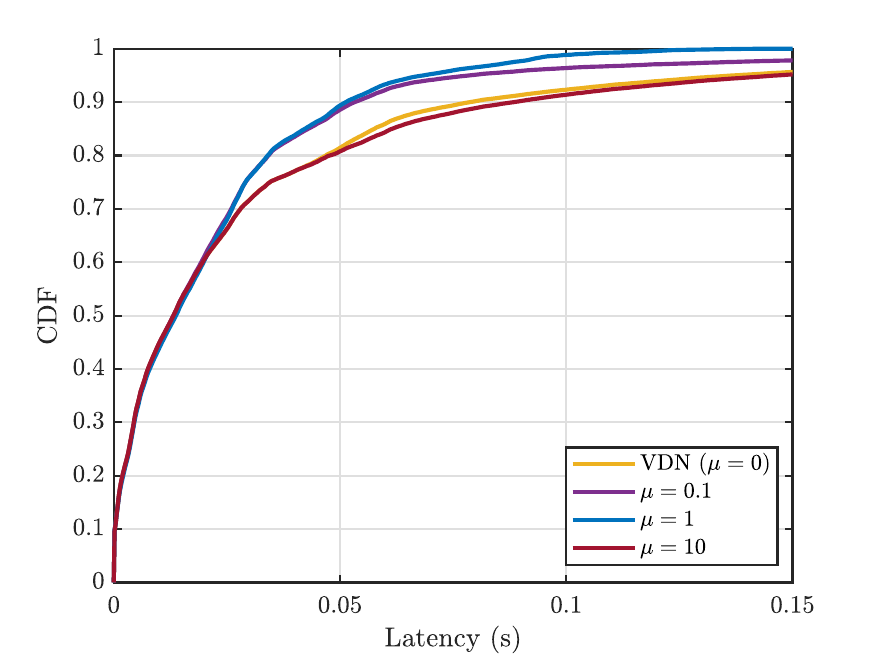}}
    %\hskip -2.8ex
    \caption{The performance of the proposed X-CODE while sweeping: (a) CQL parameter $\alpha$, and (b) explainability strength $\mu$.}
    \label{Ablation_Parameters} %\vspace{-0mm}
\end{figure}

\subsection{Ablation studies}

In this set of experiments, we study the impact of the main components of the proposed offline learning framework through ablation studies. First, Fig.~\ref{Ablation_Parameters} examines the effect of various hyperparameters on the performance of the proposed X-CODE algorithm. In particular, we focus on the CQL regularization coefficient $\alpha$ and the XAI reward-shaping strength $\mu$. Fig.~\ref{alpha} shows that the conservative penalty has an impact on the latency distribution, where setting $\alpha=0$ reduces to the original DQN objective without any conservative term. This approach suffers from over-optimistic and OOD actions exhibiting a worse latency tail. Moderate conservative regularization values, such as $\alpha=1$, provides the best latency distribution, showing that CQL term contributes in avoiding overestimation of unsupported actions in the offline dataset. In contrast, excessively large conservatism, such as $\alpha=10$, slightly degrades performance, since the policy becomes overly constrained by the dataset and less capable of improving the learned allocation strategy.

Moreover, Fig.~\ref{mu} evaluates the impact of the XAI reward-shaping strength on the latency distribution. Setting $\mu=0$ corresponds to no XAI reward relabeling, simplifying the problem to conventional VDN baseline. Without XAI-based reward relabeling, the centralized CQL policy converges to a more conservative allocation strategy that underutilizes the available resources to avoid conflicts, resulting in higher latencies. In addition, small to moderate XAI guidance improves the latency distribution, with $\mu=1$ achieves the best latency distribution, indicating that moderate attribution-based relabeling helps the offline learner obtain a more favorable balance between resource utilization, latency, and conflict avoidance. In contrast, when setting $\mu$ to relatively large values, the XAI bonus starts to dominate the original environment reward and degrades the latency distribution.

\begin{figure}[!t]
    \centering
    \subfloat[KDE \label{kde}]{\includegraphics[width=0.6875\columnwidth]{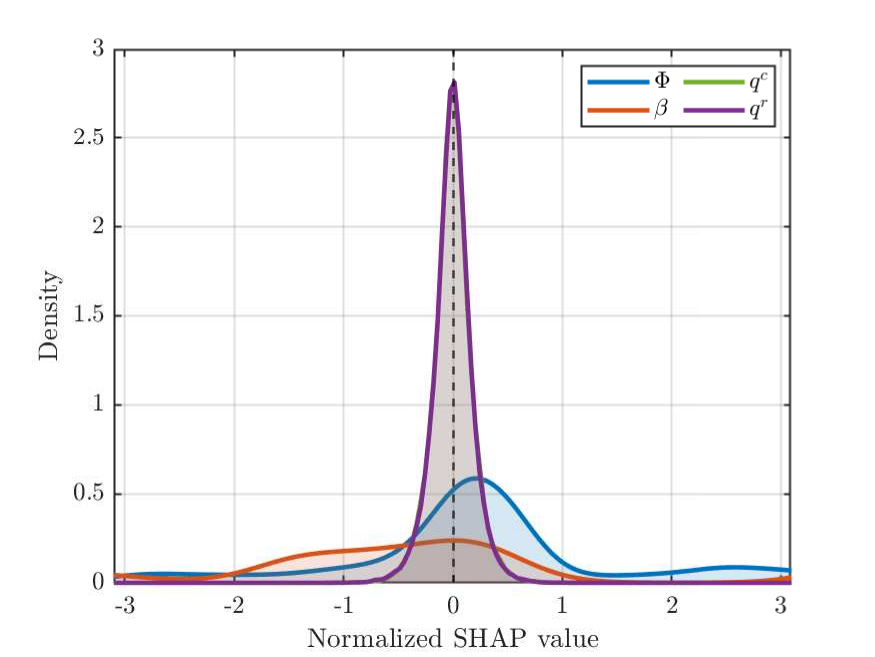}}\\
    \subfloat[Features contribution \label{feature_contrib}]{\includegraphics[width=0.6875\columnwidth]{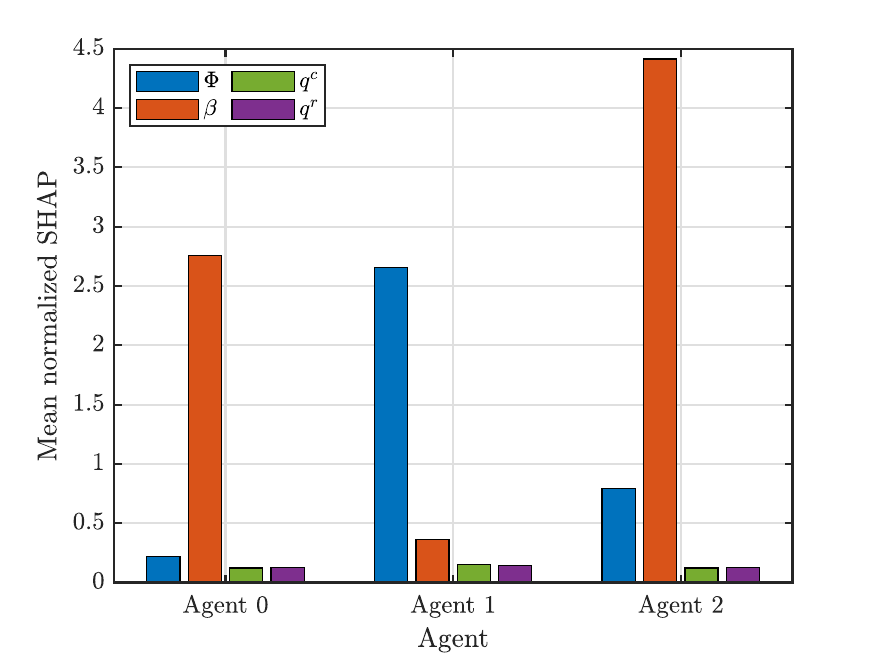}}
    %\hskip -2.8ex
    \caption{The attribution of each feature to the predictions represented in terms of the KDE plot and SHAP values.}
    \label{KDE_SHAP} %\vspace{-0mm}
\end{figure}

\subsection{XAI Insights}

Next, we analyze the proposed X-CODE framework from an explainability perspective using SHAP attributions. The objective of this experiment is to understand which features most strongly influence the centralized value estimates underlying the learned resource-allocation policies. Fig.~\ref{KDE_SHAP} reports the SHAP distributions and the corresponding feature-level importance for the centralized VDN value function. In Fig.~\ref{kde}, the KDE curves show the distribution of normalized SHAP values for the observations (\emph{i.e.}, traffic $\Phi$, CPU-gap features $\beta$, computation queue $q^c$ and transmission queue $q^r$) of the three agents. Positive SHAP values indicate that the corresponding feature increases the estimated centralized value, while negative values indicate a decreasing contribution. The wide and shifted distributions of CPU-gap features $\beta$ and traffic $\Phi$ show that these features have stronger influence on the centralized Q-function compared to the remaining features.

%\begin{figure}[!t]
%    \centering
%    \subfloat[Communication codes \label{comm_shap}]{\includegraphics[width=0.6875\columnwidth]{Figures/Fig8a_CommunicationImportance.eps}}\\
%    \subfloat[Agent contribution \label{agent_shap}]{\includegraphics[width=0.6875\columnwidth]{Figures/Fig8b_AgentContribution.eps}}
    %\hskip -2.8ex
%    \caption{An analogy between the contribution of communication codes to the predictions of the online DQN (comm + IB) model and the contribution of the agents' features to the centralized Q-function of the proposed X-CODE model.}
%    \label{Analogy} %\vspace{-0mm}
%\end{figure}

Additionally, we further confirm and demonstrate the contribution of each feature in Fig.~\ref{feature_contrib}, which reports the mean normalized SHAP contribution of each feature. The CPU-gap features of agents $1$ and $3$ ($\beta_1$ and $\beta_3$) dominate the centralized value estimate, followed by the traffic features ($\Phi_1$ and $\Phi_3$). In contrast, traffic $\Phi_{2}$ of agent $2$ dominates the contribution of agent $2$ in the centralized critic. Meanwhile, the queue-related features $q^c$ and $q^r$ are concentrated around zero, indicating that they usually provide small corrections to the critic output. This highlights that the proposed X-CODE learns slice-dependent valuation patterns, where the agents with more influential traffic or resource-demanding patterns shape the centralized allocation policies

%In this experiment, we present an analogy between explicit agents coordination using online communication codes and the proposed method's implicit agents coordination. Fig.~\ref{Analogy} provides an interpretability-based comparison between the VDN framework and the communication-based online DQN, \emph{i.e.}, online DQN (comm + IB). In Fig.~\ref{comm_shap}, we report the SHAP importance of the transmitted communication messages in the online DQN (Comm + IB) model. The SHAP importance demonstrates that messages transmitted by agents $1$ and $2$ have higher importance than those transmitted by Agent $0$. In a similar manner, Fig.~\ref{agent_shap} reports the contribution of each agent's local features to the centralized VDN value function. Similar to the DQN (Comm + IB) case, agents $1$ and $2$ have significantly higher contributions than agent $0$, showing that the centralized critic attributes most of the joint value estimation to these agents.

\begin{figure}[!t]
    \centering
    \subfloat[Signaling overhead \label{signaling}]{\includegraphics[width=0.6875\columnwidth]{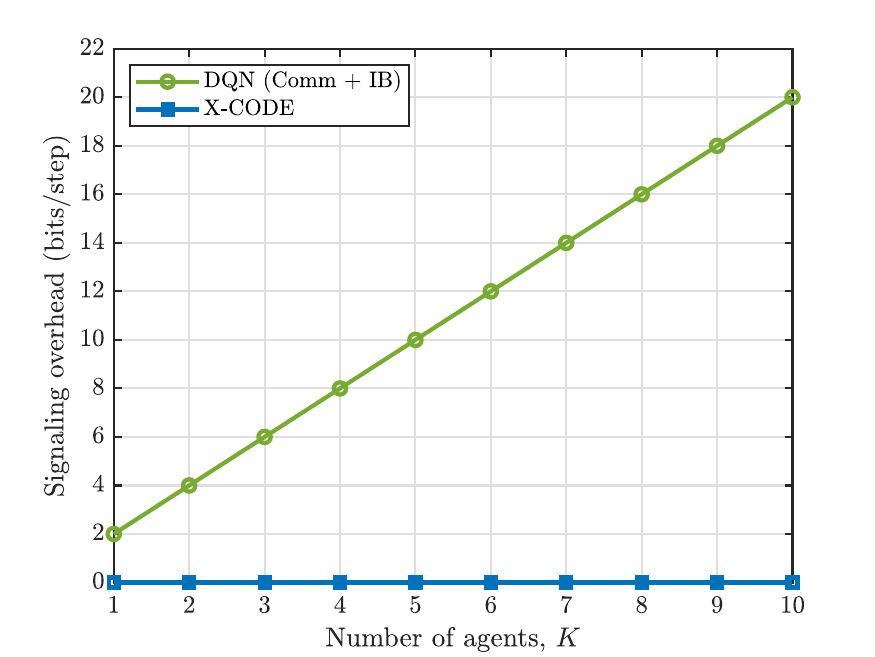}}\\
    \subfloat[Inference time \label{inference}]{\includegraphics[width=0.6875\columnwidth]{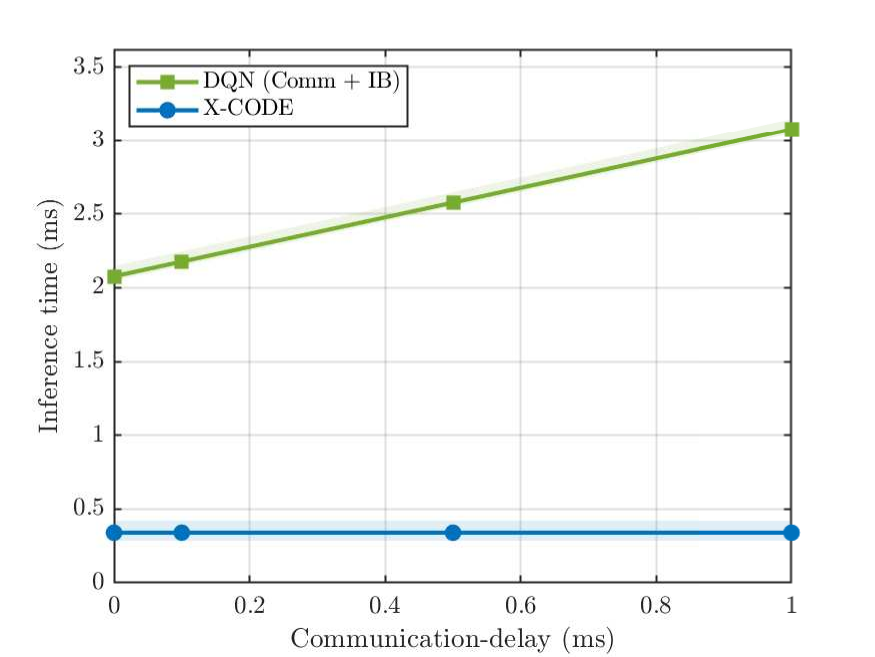}}
    %\hskip -2.8ex
    \caption{The complexity analysis of the proposed X-CODE compared to the DQN (Comm + IB) in terms of the signaling overhead and the inference time under the considered delay model.}
    \label{Complexity} %\vspace{-0mm}
\end{figure}

%This analogy highlights the different ways in which coordination is captured by the two learning methods. In the online DQN (Comm + IB) approach, coordination is explicit and appears through the learned communication messages exchanged during decentralized execution. In contrast, cross-agent dependencies in X-CODE are captured during centralized training through the shared value-decomposition objective. The SHAP analysis then reveals how strongly each agent's local features influence the resulting centralized value estimates. The main advantage of the proposed X-CODE over DQN (Comm + IB) is the mitigation of explicit communication codes during execution. The similar importance ranking observed across the two models provides a qualitative comparison rather than evidence that SHAP attributions and communication messages represent equivalent coordination mechanisms.

\subsection{Complexity Analysis}
In the final experiment in Fig.~\ref{Complexity}, we highlight the complexity analysis of the proposed method compared to the online baseline in terms of the signaling overhead and the inference time. The online baseline, DQN (Comm + IB) relies on explicit communication codes during execution, while the proposed X-CODE extracts decentralized policies without inter-agent signaling. In Fig.~\ref{signaling}, we measure the amount of signaling overhead in bits / step as function of the number of agents. In DQN (Comm + IB), each agent broadcasts one communication code per decision step, resulting in an overhead of $K \left\lceil \log_2 M \right\rceil$ bits per step, with $M = 3$ denoting the number of communication codes. In contrast, the proposed X-CODE avoids exchanging communication codes during inference, yielding zero signaling regardless the number of agents. While other offline CTDE MARL methods (\emph{e.g.}, VDN and QMIX) exhibit zero signaling, they underperform X-CODE in the achievable latency.

Fig.~\ref{inference} compares the inference time of the proposed X-CODE method and the online DQN (Comm + IB) baseline under different considered communication delays. We show the mean inference time with the shaded regions denoting the $5$th--$95$th percentile range. Since DQN (Comm + IB) requires exchanging codes during execution, its effective inference time heavily depends on the communication delay and the policy extraction time. In contrast, the proposed X-CODE only depends on the policy extraction time. Across the considered communication-delay values, X-CODE reduces the effective inference time by $88 \%$ on average compared to DQN (Comm + IB) under the evaluated delay model.

Although X-CODE introduces additional offline complexity due to conservative value learning and SHAP-based reward relabeling, this cost is incurred only during training. During inference, each agent independently evaluates its local Q-function using its local observation. These results highlight that X-CODE shifts the additional complexity to the offline training and reward-relabeling stages, while maintaining low-overhead decentralized execution.

\section{Conclusions}\label{sec:conclusions} %\vspace{1mm}

This paper investigated resource allocation in edge-enabled network slicing through resource-coupled cooperative MARL with competing slice demands. We considered multiple slices competing for shared edge resources while coordinating to avoid conflicts and maintain efficient utilization. To address the challenges imposed by limited dataset coverage and the lack of communication among agents, we proposed XAI-guided conservative decentralized execution (X-CODE). We introduced explainability-aware reward shaping to modify the relative preference among joint offline transitions using attribution information extracted from pretrained centralized critics. Under the evaluated settings, the resulting policies achieved improved resource utilization and latency while avoiding observed conflicts without inter-agent communication during execution. Extensive evaluations and ablation studies further highlighted the benefits of the proposed method in terms of effective resource utilization, low inference time and limited signaling overhead. Future work will investigate offline-to-online fine-tuning of pretrained policies and larger-scale network slicing scenarios.

\bibliographystyle{IEEEtran}
\bibliography{IEEEabrv,references}
\end{document}